\documentclass[twocolumn,prl,aps,nofootinbib,superscriptaddress]{revtex4}
\usepackage{graphicx}
\usepackage{amsmath}
\begin{document}

\newcommand \be{\begin{equation}}
\newcommand \bea{\begin{eqnarray}}
\newcommand \ee{\end{equation}}
\newcommand \eea{\end{eqnarray}}
\newcommand{\rar}{\rightarrow}
\newcommand{\eq}{equation}
\newcommand{\rp}{\right)}
\newcommand{\lp}{\left(}
\newcommand{\E}{{\rm E}}


\title{Collective Origin of the Coexistence of Apparent RMT Noise\\ and Factors
in Large Sample Correlation Matrices}

\author{Y. Malevergne}
\affiliation{Laboratoire de Physique de la Mati\`ere Condens\'ee CNRS UMR 6622\\
Universit\'e de Nice-Sophia Antipolis, 06108 Nice Cedex 2, France}
\affiliation{ Institut de Science Financi\`ere et d'Assurances -
Universit\'e Lyon I\\ 43, Bd du 11 Novembre 1918, 69622 Villeurbanne Cedex,
France}

\author{D. Sornette}
\affiliation{Laboratoire de Physique de la Mati\`ere Condens\'ee CNRS UMR 6622\\
Universit\'e de Nice-Sophia Antipolis, 06108 Nice Cedex 2, France}
\affiliation{Institute of Geophysics and Planetary Physics
and Department of Earth and Space Science\\
University of California, Los Angeles, California 90095, USA}

\begin{abstract}
Through simple analytical calculations
and numerical simulations, we demonstrate the generic
existence of a self-organized macroscopic state in any large 
multivariate system
possessing non-vanishing average correlations between a finite
fraction of all pairs of elements.
The coexistence of an eigenvalue spectrum predicted by random matrix theory (RMT)
and a few very large eigenvalues in large empirical correlation matrices
is shown to result from a bottom-up collective effect of the underlying time series
rather than a top-down impact of factors. Our results, in excellent
agreement with previous results obtained on large financial correlation matrices,
show that there is relevant information also in the bulk of the eigenvalue
spectrum and rationalize the presence of market factors previously 
introduced in an {\it ad hoc} manner.

\end{abstract}

\maketitle

\vskip 0.5cm
\vskip 0.5cm


Since Wigner's seminal idea to apply random matrix theory (RMT) to interpret
the complex spectrum of energy levels in nuclear physics \cite{Wigner}, RMT
has made enormous progress \cite{Mehta} with many applications in physical
sciences and elsewhere such as in meteorology \cite{SP01} and 
image processing \cite{SM}. A new application was
proposed a few years ago to the problem of correlations
between financial assets and to the portfolio optimization problem.
It was shown that, among the eigenvalues and principal
components of the empirical
correlation matrix of the returns of hundreds of
asset on the New York Stock Exchange
(NYSE), apart from the few highest eigenvalues, the
marginal distribution of the other eigenvalues and eigenvectors
closely resembles the spectral
distribution of a positive symmetric random matrix with maximum entropy,
suggesting that the correlation matrix does not contain 
any specific information beyond these few largest eigenvalues
and eigenvectors \cite{Laloux}. These results apparently invalidate the
standard mean-variance portfolio optimization theory \cite{Marco}
consecrated by the financial industry \cite{riskmetrics}
and seemingly support the rationale
behind factor models such as the capital asset pricing model (CAPM) \cite{CAPM} and 
the arbitrage pricing theory (APT) \cite{APT}, where the
correlations between a large number of assets are represented
through a small number of so-called market factors. 
Indeed, if the spectrum of eigenvalues of the empirical covariance or
correlation matrices are predicted by RMT,
it seems natural to conclude that
there is no usable information in these matrices
and that empirical covariance matrices should not be used for
portfolio optimization. In contrast, if one detects deviations
between the universal -- and therefore non-informative -- part
of the spectral properties of empirically estimated covariance and correlation
matrices and those of the relevant ensemble of random matrices, this
may quantify the amount of real information that can be used in portfolio
optimization from the ``noise'' that should be discarded.

More generally, in many different
scientific fields, one needs to determine the nature and amount of information
contained in large covariance
and correlation matrices. This occurs as soon as one attempts to estimate
very large covariance and correlation matrices in
multivariate dynamics of systems exhibiting
non-Gaussian fluctuations with fat tails and/or long-range time
correlations with intermittency. In such cases, the convergence
of the estimators of the large covariance and correlation matrices is often
too slow for all practical purposes. 
The problem becomes even more complex with
time-varying variances and covariances as occurs 
in systems with heteroskedasticity \cite{Engle} or with regime-switching
\cite{regimeswitch}. A prominent example where such difficulties arise is the
data-assimilation problem in engineering and in meteorology where forecasting
is combined with observations iteratively through the
Kalman filter, based on the estimation and
forward prediction of large covariance matrices \cite{Kalman}.

As we said in the context of financial time series, the rescuing
strategy is to invoke the existence of a
few dominant factors, such as an overall market factor
and the factors related to firm size, firm industry and book-to-market equity,
thought to embody most of the 
relevant dependence structure between the studied time series \cite{fama}. Indeed,
there is no doubt that observed equity prices respond to a wide
variety of unanticipated factors, but there is much weaker evidence that
expected returns are higher for equities that are more sensitive to these
factors, as required by Markowitz's mean-variance theory, 
by the CAPM and the APT \cite{Roll}. This severe failure of
the most fundamental finance theories could
conceivably be attributable to an inappropriate proxy for the market
portfolio, but nobody has been able to show that this is really the correct
explanation. This remark constitutes the crux of the problem:  
the factors invoked to model the cross-sectional dependence between assets
are not known in general and are either postulated based on economic intuition in 
financial studies or obtained as black box results in the recent 
analyses using RMT \cite{Laloux}.

Here, we show that the existence of factors
results from a collective effect of the assets,
similar to the emergence of a macroscopic self-organization of
interacting microscopic constituents. For this, 
we unravel the general physical origin of the large eigenvalues 
of large covariance and correlation matrices and provide a complete
understanding of the coexistence of features resembling
properties of random matrices and of large ``anomalous'' eigenvalues.
Through simple analytical calculations
and numerical simulations, we demonstrate the generic
existence of a self-organized macroscopic state in any large system
possessing non-vanishing average correlations between a finite
fraction of all pairs of elements.

Let us first consider a large system of size $N$ with correlation
matrix $C$ in which every non-diagonal pairs of elements exhibits the
same correlation coefficient $C_{ij}=\rho$ for $i \neq j$ and $C_{ii}=1$.
Its eigenvalues are 
\be
\lambda_{1} =1+(N-1)\rho ~~{\rm and} ~~\lambda_{i \ge 2}=1-\rho
\label{resul}
\ee
with multiplicity
$N-1$ and with $\rho \in (0,1)$ in order for the correlation matrix to remain
positive definite. Thus, in the 
thermodynamics limit $N \to \infty$, even for a weak positive correlation $\rho 
\to 0$ (with $\rho N \gg 1$), 
a very large eigenvalue appears, associated with the delocalized
eigenvector $v_1 =(1/\sqrt{N}) (1, 1, \cdots,1)$, which dominates
completely the correlation structure of the system. This trivial example
stresses that the
key point for the emergence of a large eigenvalue is not the strength
of the correlations, provided that they do not vanish, but the large size $N$ of
the system.

This result (\ref{resul}) still holds qualitatively when the
correlation coefficients are all distinct. To see this, it is 
convenient to use a perturbation approach. We thus
add a small random component to each correlation coefficient:
\be
C_{ij} = \rho + \epsilon \cdot a_{ij}  ~~~{\rm for}~~i \neq j~,
\label{noise}
\ee
 where the coefficients $a_{ij}=a_{ji}$ have zero
mean, variance $\sigma^2$ and are independently distributed (There are
additional constraints on the 
support of the distribution of the $a_{ij}$'s in order for the
matrix $C_{ij}$ to remain positive definite with probability one). The determination
of the eigenvalues and eigenfunctions of $C_{ij}$ is performed using the
perturbation theory developed in quantum mechanics \cite{cohen} up to the second order in
$\epsilon$. We find that the largest eigenvalue becomes
\be
\E[\lambda_{1}]=(N-1)\rho +1 + \frac{(N-1)(N-2)}{N^2} \cdot \frac{\epsilon^2
\sigma^2}{\rho} + {\cal O}(\epsilon^3)
\label{resul2}
\ee
while, at the same order, the corresponding eigenvector $v_1$ remains
unchanged. The degeneracy of the eigenvalue $\lambda=1-\rho$ is broken
and leads to a complex set of smaller eigenvalues described below. 

In fact, this result (\ref{resul2}) can be generalized to
the non-perturbative domain of any correlation 
matrix with independent random coefficients $C_{ij}$, provided that
they have the same mean value $\rho$ and variance $\sigma^2$. Indeed, it has
been shown \cite{FK81} that, in such a case, the expectations of the largest and
second largest eigenvalues are
\bea
\E[\lambda_1] &=& (N-1)\cdot \rho +1 + \sigma^2/\rho + {\it o}(1)~, \label{eq:l1}\\
\E[\lambda_2] &\le&  2 \sigma \sqrt{N} + {\cal O}(N^{1/3} \log N)~. \label{eq:l22}
\eea
Moreover, the statistical fluctuations of these two largest eigenvalues are
asymptotically (for large fluctuations $t > {\cal O}(\sqrt{N})$) 
bounded by a Gaussian distribution according to the following 
large deviation theorem
\be
\Pr\{ | \lambda_{1,2} - \E[\lambda_{1,2}] | \ge t \} \le e^{-c_{1,2}t^2}~,
\ee
for some positive constant $c_{1,2}$ \cite{KV00}.

\begin{figure}
\begin{center}
\includegraphics[width=8.5cm]{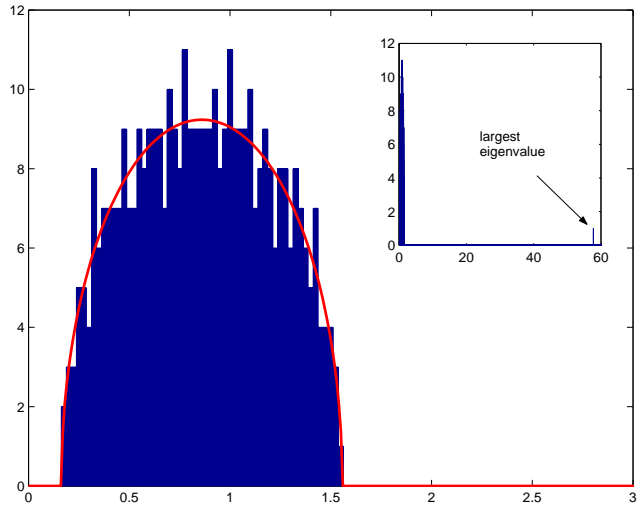}
\end{center}
\caption{\label{fig:1} Spectrum of eigenvalues of a random correlation matrix with
average correlation coefficient
$\rho=0.14$ and standard deviation of the correlation
coefficients $\sigma = 0.345/\sqrt{N}$. The size $N=406$ of the matrix
is the same as in previous studies \cite{Laloux} for the sake of comparison.
The continuous curve is
the theoretical translated semi-circle distribution of eigenvalues
describing the bulk
of the distribution which passes the Kolmogorov test. The center value $\lambda= 1- \rho$
ensures the conservation of the trace equal to $N$.
There is no adjustable parameter. The inset represents
the whole spectrum with the largest eigenvalue whose size is in agreement
with the prediction $\rho N =56.8$.}
\end{figure}

This result is very different from that obtained when the mean value $\rho$
vanishes. In such a case, the distribution of eigenvalues of the random
matrix $C$ is given by the semi-circle law \cite{Mehta}. However, 
due to the presence of the
ones on the main diagonal of the correlation matrix $C$, the center of the circle is
not at the origin but at the point $\lambda=1$. Thus, the distribution of the
eigenvalues of random correlation matrices
with zero mean correlation coefficients
is a semi-circle of radius $2 \sigma \sqrt{N}$ centered at $\lambda =1$.

The result (\ref{eq:l1}) is deeply related to the so-called ``friendship theorem''
in mathematical graph theory, which states that, in any finite graph such that any two
vertices have exactly one common neighbor, there is one and only one vertex
adjacent to all other vertices \cite{Erdos}. A more heuristic but equivalent 
statement is that, in a group of people such that any pair of persons have exactly
one common friend, there is always one person (the ``politician'') who is
the friend of everybody. The connection is established by taking the non-diagonal
entries $C_{ij}$ ($i \neq j$) equal to Bernouilli random variable with parameter $\rho$,
that is, $Pr[C_{ij}=1]=\rho$ and $Pr[C_{ij}=0]=1-\rho$.
Then, the matrix $C_{ij}-I$, where $I$
is the unit matrix, becomes nothing but the adjacency matrix of the random graph $G(N, \rho)$
\cite{KV00}. The proof of \cite{Erdos} of the ``friendship theorem''
indeed relies on the $N$-dependence of the largest
eigenvalue and on the $\sqrt{N}$-dependence of the second largest eigenvalue of $C_{ij}$
as given by (\ref{eq:l1}) and (\ref{eq:l22}).

Figure~\ref{fig:1} shows the distribution of eigenvalues of a random
correlation matrix.
The inset shows the largest eigenvalue lying at the predicting size $\rho N =56.8$,
while the bulk of the eigenvalues are much smaller
and are described by a modified semi-circle law centered 
on $\lambda= 1- \rho$, in the limit of large $N$.  
The result on the largest eigenvalue emerging
from the collective effect of the cross-correlation between all $N (N-1)/2$ pairs
provides a novel perspective
to the observation \cite{Rollcrash} that the only reasonable explanation
for the simultaneous crash of 23 stock markets worldwide in October 1987 is
the impact of a world market factor: according to our demonstration, the
simultaneous occurrence of
significant correlations between the markets worldwide is bound to 
lead to the existence
of an extremely large eigenvalue, the world market factor constructed by ...
a linear combination of the 23 stock markets! What our result shows is that 
invoking factors to explain the cross-sectional structure of stock returns
is cursed by the chicken-and-egg problem: factors exist because stocks
are correlated; stocks are correlated because of common factors impacting them.

\begin{figure}
\begin{center}
\includegraphics[width=8.5cm]{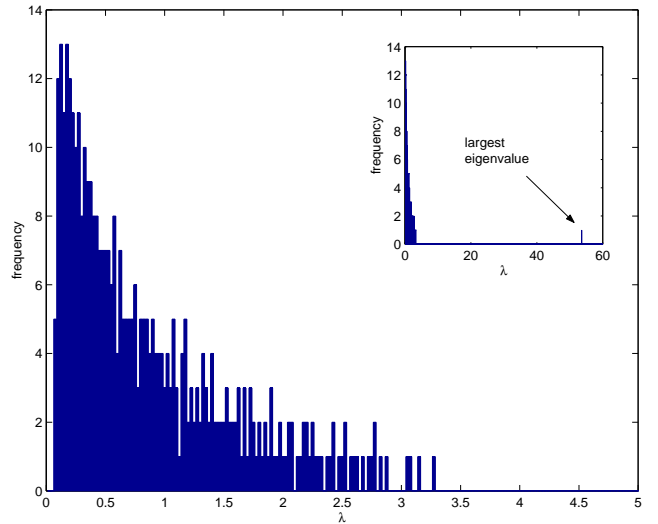}
\end{center}
\caption{\label{fig:2} Spectrum estimated from the sample correlation matrix
obtained from $N=406$ time series of length $T=1309$ (the same length as in
 \cite{Laloux}) with the same
theoretical correlation matrix as that presented in figure~\ref{fig:1}.}
\end{figure}

Figure~\ref{fig:2} shows the eigenvalues distribution
of the sample correlation matrix reconstructed by sampling $N=406$
time series of length $T=1309$ generated with a given correlation matrix $C$
with theoretical spectrum shown in figure~\ref{fig:1}.
The largest eigenvalue is again very close to the prediction $\rho N =56.8$
while the bulk of the distribution departs very strongly from the 
semi-circle law and is not far from the Wishart prediction, as expected
from the definition of the Wishart ensemble as 
the ensemble of sample covariance matrices of 
Gaussian distributed time series with unit variance and zero mean. 
A Kolmogorov test shows however that the bulk of the spectrum is not 
in the Wishart class, in contradiction with previous claims lacking
formal statistical tests \cite{Laloux}. This result holds 
for different
simulations of the sample correlation matrix and different realizations of the
theoretical correlation matrix with the same parameters ($\rho$, $\sigma$). 
The statistically significant departure from the Wishart prediction
implies that there is actually some information in the bulk of the
spectrum of eigenvalues, which is intimately coupled with the existence
of the largest eigenvalue. We have also checked that these results
remain robust for non-Gaussian distribution of returns as long as
the second moments exist. Indeed, correlated time series with multivariate
Gaussian or Student distributions with three degrees of freedom
(which provide more acceptable proxies for financial time series 
\cite{Gopikrishnan}) give no discernible differences in the spectrum
of eigenvalues. This is surprising as the estimator of a correlation
coefficient is asymptotically Gaussian for time series with finite
fourth moment and L\'evy stable otherwise \cite{MS2}. 

Up to now, we have focused on the collective mechanism at the origin
of the very large eigenvalue of order $N$. Empirically \cite{Laloux}, a few 
other eigenvalues have an amplitude of the order of $5-10$ 
that deviate significantly from the bulk of the distribution. These eigenvalues
cannot be obtained by a matrix of the form (\ref{noise}) with identically
independently distributed coefficients $a_{ij}$'s.
Our analysis provides a very simple constructive mechanism for them.
The solution consists in considering, as a first approximation, the block
diagonal matrix $C'$ with diagonal elements made
of the matrices $A_1, \cdots, A_p$ of sizes $N_1, \cdots, N_p$ with
$\sum N_i = N$, constructed according to (\ref{noise}) such that 
each matrix $A_i$ has the average correlation coefficient
$\rho_i$. When the coefficients of the matrix $C'$
outside the matrices $A_i$ are zero, the spectrum of $C'$ is given by the
union of all the spectra of the $A_i$'s, which are each dominated by a large
eigenvalue $\lambda_{1,i} \simeq  \rho_i \cdot N_i$. The spectrum of $C'$ then
exhibits $p$ large eigenvalues. 

\begin{figure}
\begin{center}
\includegraphics[width=8.5cm]{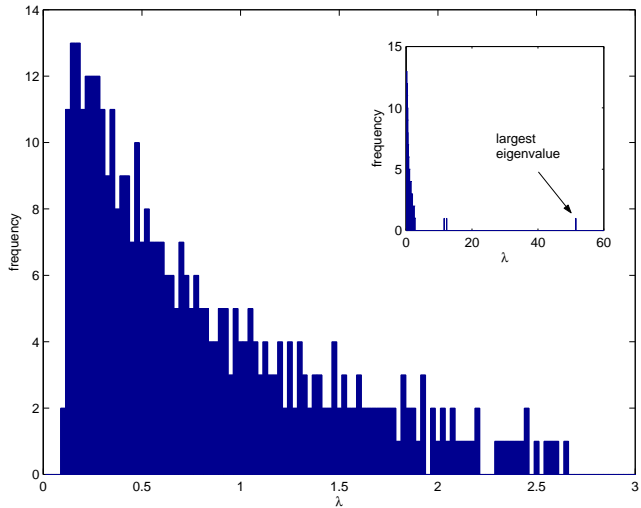}
\end{center}
\caption{\label{fig:3} Spectrum of eigenvalues estimated from the sample correlation matrix
of $N=406$ time series of length $T=1309$. The times series have been constructed
from a multivariate Gaussian distribution with a correlation matrix
made of three block-diagonal matrices of sizes respectively equal to $130$, $140$ and $136$
and mean correlation coefficients equal to $0.18$ for all of them. The off-diagonal
elements are all equal to $0.1$. The same results hold if the off-diagonal 
elements are random. }
\end{figure}

Each block $A_i$ can be interpreted as a sector of the economy,
including all the companies belonging to a same industrial branch and the
eigenvector associated with each
largest eigenvalue represents the main factor driving this sector of activity
\cite{Mantegnapath}. For similar sector sizes $N_i$ and average correlation
coefficients $\rho_i$, the largest eigenvalues are of the same
order of magnitude. In order to recover a very large unique eigenvalue,
we reintroduce some coupling constants outside the block diagonal
matrices. A well-known result of perturbation theory in quantum mechanics
states that such coupling
leads to a repulsion between the eigenstates, which can be observed
in figure~\ref{fig:3} where $C'$ has been constructed with three block matrices
$A_1$, $A_2$ and $A_3$ and non-zero
 off-diagonal coupling described in the figure caption.
These values allow us to quantitatively
replicate the empirical finding of Laloux {\it et al.} in \cite{Laloux}, where
the three first eigenvalues are approximately $\lambda_1 \simeq 57$, $\lambda_2
\simeq 10$ and $\lambda_3 \simeq 8$. The bulk of the spectrum (which excludes
the three largest eigenvalues) is similar to the Wishart distribution but again
statistically different from it as tested with a Kolmogorov test. There
is thus significant deviation from the predictions of RMT not only for the 
largest eigenvalues but also in the bulk.

As a final remark, expressions (\ref{resul2},\ref{eq:l1}) and our numerical tests
for a large variety of correlation matrices show that the
delocalized eigenvector  $v_1 =(1/\sqrt{N}) (1, 1, \cdots,1)$,
associated with the largest eigenvalue is extremely robust
and remains (on average) the same for any large system. Thus, even for
time-varying correlation matrices -- as in finance with important 
heteroskedastic effects -- the composition of the main factor remains almost the same. This
can be seen as a generalized limit theorem reflecting the bottom-up
organization of broadly correlated time series.

This work was partially supported by
the James S. Mc Donnell Foundation 21st century scientist award/studying
complex system.


\vskip -0.5cm
\begin{thebibliography}{99}

\bibitem{Wigner} Wigner, E.P., Ann. Math. {\bf 53}, 36 (1951).

\bibitem{Mehta}  Mehta, M.L., Random matrices, 2nd ed. (Boston:
 Academic Press, 1991).

\bibitem{SP01} Santhanam, M.S. and P. K. Patra, Phys. Rev. E {\bf 64}, 016102 (2001).

\bibitem{SM} Setpunga, A.M. and P.P. Mitra,  Phys. Rev E {\bf 60}, 3389 (1999).

\bibitem{Laloux} Laloux, L. et al.,
Phys. Rev. Lett. {\bf 83}, 1467 (1999);
Plerou, V., et al., Phys. Rev. Lett. {\bf 83}, 1471 (1999); Phys. Rev. E; 
Maslov, S., Physica A {\bf 301}, 397 (2001); Plerou, V. et al.,
Phys. Rev E {\bf 6506} 066126 (2002).

\bibitem{Marco} Markowitz, H., Portfolio selection: Efficient diversification of
investments (John Wiley and Sons, New York, 1959).

\bibitem{riskmetrics} RiskMetrics Group, RiskMetrics (Technical Document, NewYork: J.P.
Morgan/Reuters, 1996).

\bibitem{CAPM}  Sharpe, W.F., J. Finance (September), 425 (1964); 
 Lintner, J., Rev. Econ. Stat. (February), 13 (1965);
  Mossin, J., Econometrica (October), 768 (1966);
   Black, F., J. Business (July), 444 (1972).
  
\bibitem{APT}  Ross, S.A., J. Economic Theory (December), 341 (1976).

\bibitem{Engle} Engle, R.F. and K. Sheppard, NBER Working Paper No. W8554 (2001).

\bibitem{regimeswitch} Schaller, H. and van Norden, S.,
Appl. Financial Econ. {\bf 7}, 177 (1997).

\bibitem{Kalman} Brammer, K.,
Kalman-Bucy filters (Gerhard Siffling,  Norwood, MA: Artech House, 1989).

\bibitem{fama} Fama, E.F. and Kenneth R., 
J. Finance {\bf 51}, 55 (1996); 
J. Financial Econ. {\bf 33}, 3 (1993); Fama, E.F. et al., 
Financial Analysts J. {\bf 49}, 37 (1993).
     
\bibitem{Roll} Roll, R., Financial Management {\bf 23}, 69 (1994).

\bibitem{cohen} Cohen-Tannoudji, C., B. Diu and F. Laloe,
Quantum mechanics (New York: Wiley, 1977).

\bibitem{FK81} F\"redi Z. and J. Koml\'os, Combinatorica 1, 233-241 (1981).

\bibitem{KV00} Krivelevich, M. and V. H. Vu, math-ph/0009032 (2000).

\bibitem{Erdos} Erdos, P. et al., Studia Sci. Math. 1, 215 (1966).

\bibitem{Rollcrash} R. Roll, Financial Analysts J. {\bf 44}, 19 (1988).

\bibitem{Gopikrishnan}
Gopikrishnan, P. et al., Eur. Phys. Journal B {\b 3}, 139 (1998); 
Guillaume, D.M., et al., Finance and Stochastics {\bf 1}, 95 (1997);
Lux, L., Appl. Financial Economics {\bf 6}, 463 (1996);
Pagan, A., J. Emp. Fin. {\bf 3}, 15 (1996).

\bibitem{MS2} Davis, R.A. and J.E. Marengo, Commun. Statist.-Stochastic Models {\bf 6},
483 (1990); Meerschaert, M.M. and H.P. Scheffler,
J. Time Series Anal. {\bf 22}, 481 (2001).

\bibitem{Mantegnapath} Mantegna, R.N., Eur. Phys. J. {\bf 11}, 193 (1999);
Marsili, M., Quant. Fin. {\bf 2}, 297 (2002).

\end{thebibliography}
\end{document}